\documentclass[11pt,a4paper,aps,preprintnumbers,superscriptaddress,amssymb,nofootinbib]{revtex4}
\usepackage[english]{babel}
\usepackage{epsfig}  
\usepackage{graphicx}
\usepackage{hyperref}
\hypersetup{
    colorlinks=true,
    citecolor=red,
    linkcolor=blue,
    filecolor=green,      
    urlcolor=magenta,
}
\usepackage{color}
\usepackage{float}
\usepackage{amsfonts}
\usepackage{amsmath}
\usepackage{slashed}
\usepackage{soul}
\large

\begin{document}

\title{New physics in $b\rightarrow se^+e^-$: A model independent analysis}

\author{Ashutosh Kumar Alok}
\email{akalok@iitj.ac.in}
\affiliation{Indian Institute of Technology Jodhpur, Jodhpur 342037, India}

\author{Suman Kumbhakar}
\email{ksuman@iisc.ac.in}
\affiliation{Centre for High Energy Physics, Indian Institute of Science, Bangalore 560012, India }

\author{Jyoti Saini}
\email{saini.1@iitj.ac.in}
\affiliation{Indian Institute of Technology Jodhpur, Jodhpur 342037, India}

\author{S. Uma  Sankar}
\email{uma@phy.iitb.ac.in}
\affiliation{Indian Institute of Technology Bombay, Mumbai 400076, India}

\begin{abstract}
The lepton universality violating flavor ratios $R_K/R_{K^*}$ indicate new physics either in $b \to s \mu^+ \mu^-$ or in
$b \to s e^+ e^-$ or in both. If the new physics is only $b \to s e^+ e^-$ transition, the corresponding new physics operators, in principle, can have any Lorentz structure. In this work, we perform a model independent analysis of new physics only in  $b \to se^+e^-$  decay by considering effective operators either one at a time or two similar operators at a time. We include all the measurements in $b\rightarrow se^+e^-$ sector along with $R_K/R_{K^*}$ in our analysis. We show that various new physics scenarios with vector/axial-vector operators can account for $R_K/R_{K^*}$ data but those with scalar/pseudoscalar operators and with tensor operators can not. We also show that the azimuthal angular observable $P_1$ in $B \to K^* e^+ e^-$ decay is most suited to discriminate between the different allowed solutions.
\end{abstract}

\maketitle 
\newpage

\section{Introduction} 
The current measurements in $b \to s\ell^+ \ell^-$ ($\ell=e,\,\mu$) sector show some significant tensions with the predictions of the Standard Model (SM). These include measurements of the lepton flavor universality (LFU) violating ratios $R_{K}$ and $R_{K^*}$ by the LHCb collaboration. In 2014, the LHCb collaboration reported the first measurement of the ratio $R_K \equiv  \Gamma(B^+ \to K^+\mu^+\mu^-)/\Gamma(B^+ \to K^+e^+e^-)$ in the di-lepton invariant mass-squared, $q^2$, range $1.1 \le q^2 \le 6.0$ GeV$^2$ \cite{rk}. The measured value $0.745^{+0.090}_{-0.074}(\rm stat.)\pm 0.036(\rm syst.)$  deviates from the SM prediction of $\approx 1$ \cite{Hiller:2003js,Bordone:2016gaq,Bouchard:2013mia} by 2.6$\sigma$ \footnote{A rigorous analysis of QED corrections in $R_K$ has been recently performed in \cite{Isidori:2020acz}. }. Including the Run-II data and an update of the Run-I analysis, the value of  $R_{K}$ was updated in Moriond-2019.  The updated value $0.846^{+0.060}_{-0.054}\,(\rm stat.)^{+0.016}_{-0.014}({\rm syst.})$~\cite{Aaij:2019wad} is  still $\simeq 2.5\sigma$ away from the SM. 
  
The hint of LFU violation is further observed in another flavor ratio $R_{K^*}$. This ratio $R_{K^*} \equiv \Gamma (B^0 \to K^{*0} \mu^+\mu^-)/\Gamma(B^0 \to K^{*0} e^+ e^-)$ was measured  in the low ($0.045 \leq q^2 \leq 1.1$  GeV$^2$) as well as in the central ($1.1 \leq q^2 \leq 6.0$ GeV$^2$) $q^2$ bins by the LHCb collaboration~\cite{Aaij:2017vbb}. The measured values are $0.660^{+0.110}_{-0.070}(\rm stat.)\pm 0.024(\rm syst.)$ for the low $q^2$ bin and $0.685^{+0.113}_{-0.069}(\rm stat.)\pm 0.047(\rm syst.)$  for the central $q^2$ bin. These measurements differ from the SM predictions of $R_{K^*}^{\rm low} = 0.906\pm 0.028$ and $R_{K^*}^{\rm central} = 1.00\pm 0.01$~\cite{Bordone:2016gaq}  by $\sim 2.5\sigma$ and $\sim 3\sigma$, respectively. Later Belle collaboration announced their first results on the measurements of $R_{K^*}$ in different $q^2$ bins for both $B^0$ and $B^+$ decay modes~\cite{Abdesselam:2019wac}. The measured values suffer from large statistical uncertainties and hence consistent with the SM predictions.
The  ratios $R_{K}$ and $R_{K^*}$ are essentially free from the hadronic uncertainties, making them extremely sensitive to new physics (NP) in $b\rightarrow s e^+ e^-$ or/and $b\rightarrow s\mu^+\mu^-$ transition(s). 

Further, there are a few anomalous measurements which are related to possible NP in  $b \to s \mu^+ \mu^-$ transition only. These include measurements of angular observables, in particular $P'_5$, in $B \to K^* \, \mu^+\,\mu^-$ decay \cite{Kstarlhcb1,Kstarlhcb2,Aaij:2020nrf} and the branching ratio of $B_s \to \phi\, \mu^+\,\mu^-$ \cite{bsphilhc2}. By virtue of these measurements, it is natural to assume NP only in the muon sector to accommodate all 
$b \to s\ell^+ \ell^-$ data. A large number of global analyses of $b \to s\ell^+ \ell^-$ data have been performed under this assumption \cite{Alguero:2019ptt,Alok:2019ufo,Ciuchini:2019usw,DAmico:2017mtc,Aebischer:2019mlg,Kowalska:2019ley,Arbey:2019duh}.
NP amplitude in $b \to s \mu^+ \mu^-$ must have destructive interference with the SM amplitude to account for 
$R_K, R_{K^*} < 1$. Hence the NP operators in this sector are constrained to be in vector/axial-vector form. The global 
analyses found three different combinations of such operators which can account for all the data. 
Possible methods to distinguish between these allowed NP solutions are investigated in refs.~\cite{Kumar:2017xgl,
Kumbhakar:2018uty,Alok:2020bia,Bhutta:2020qve}.
However, the predicted value of $R_{K^*}^{\rm low}$ for the solutions with NP only in $b \to s \mu^+ \mu^-$  still 
differs significantly from the measured value. This requires presence of NP in  $b\rightarrow se^+e^-$ along with 
$b\rightarrow s\mu^+\mu^-$, see for e.g,~\cite{Kumar:2019qbv,Datta:2019zca}.

While the LFU ratios  $R_{K}$ and $R_{K^*}$ are theoretically clean, other observables 
in $b \to s \mu^+ \mu^-$ sector which show discrepancy with SM, 
in particular the angular observables $B \to K^*\mu^+\mu^-$ and $B_s \to \phi \mu^+\mu^-$, are subject to significant hadronic uncertainties dominated by undermined power corrections.  
So far, the power corrections can be estimated only in the inclusive decays. For exclusive decays, there are no theoretical description of power corrections within QCD factorization and SCET framework. The possible NP effects in these observables can be masked by
such corrections. The disagreement with the  SM depends upon the guess value of power corrections. Under the assumption of 
$\rm \sim 10\%$ non-factorisable power corrections in the SM predictions, the measurements of these observables show 
deviations from the SM at the level of 3-4$\sigma$. However, if one assumes a sizable non-factorisable power corrections, the experimental data can be accommodated within the SM itself \cite{Ciuchini:2015qxb,Hurth:2016fbr,Chobanova:2017ghn,Hurth:2020rzx}. It is therefore expected that  these tensions might stay unexplained until  Belle-II can measure the corresponding observables in the inclusive $b \to s \mu^+ \mu^-$ modes \cite{Hurth:2016fbr}.

Therefore, if one considers the discrepancies in clean observables in $b \to s\ell^+ \ell^-$ sector, which are $R_{K}$ and $R_{K^*}$, then NP only in $b\rightarrow s e^+e^-$ is as natural solution as NP in $b\rightarrow s\mu^+\mu^-$ sector. In this work, we consider this possibility and perform a model independent analysis with NP restricted to $b\rightarrow s e^+e^-$ sector \footnote{A fit to $R_K$ and $R_{K^*}$ data along with the branching ratio
of $B_s \to \mu^+ \mu^-$, by assuming NP only in the muon couplings,
was performed in \cite{Alguero:2019ptt, DAmico:2017mtc, Arbey:2019duh}. The NP couplings obtained from this fit to clean observables, lead to deviations in the predictions
in the other anomalous observables, which are in the direction
indicated by the experimental measurements.}. In this scenario, we need the NP operators to increase the denominators of $R_K$ and $R_{K^*}$. 
Hence, the need for interference with SM amplitude is no longer operative.
We consider NP in the form of vector/axial-vector (V/A), scalar/pseudoscalar (S/P) and tensor (T) operators. 
We show that solutions based on V/A operators predict values of $R_K/R_{K^*}$, including $R_{K^*}^{\rm low}$,
which are in good agreement with the measured values. 
The scalar NP operators can account for the reduction in $R_K$ but not in $R_{K^*}$ and hence are ruled out. 
The coefficients of pseudoscalar operators are very severely constrained by the current bound on the branching 
ratio of $B_s \to e^+e^-$ and these operators do not lead to a reduction of $R_K/R_{K^*}$.
It is not possible to get a solution to the $R_K/R_{K^*}$ problem using only tensor operators~\cite{Hiller:2014yaa} but a solution is
possible in the form of a combination of V/A and T operators, as shown in ref. \cite{Bardhan:2017xcc}.
In this work, we will limit ourselves to solutions involving either one NP operator or two
similar NP operators at a time. We will not consider solutions with two or more dissimilar operators.

The paper is organized as follows. In Sec.~\ref{method}, we discuss the methodology adopted in this work. 
The fit results for NP in the form of V/A  operators are shown in Sec.~\ref{VA}. In Sec.~\ref{unique}, we discuss methods to discriminate
between different V/A solutions and comment on the most effective angular observables which can achieve this
discrimination. 
Finally, we present our conclusions in Sec.~ \ref{concl}.  

\section{Methodology}
\label{method}
We analyze the $R_K/R_{K^*}$ anomalies within the framework of effective field theory (EFT) by assuming NP only in 
$b \to s e^+ e^-$ transition. We intend to identify the set of operators which can account for the measurements of 
$R_K/R_{K^*}$. We consider NP in the form of V/A, S/P and T operators and analyze scenarios with either one NP operator
(1D) at a time or two similar NP operators (2D) at a time.

In the SM, the effective Hamiltonian for $ b\to s\ell^+ \ell^- $ transition is
\begin{eqnarray}
\mathcal{H}^{\rm SM} &=& - \frac{4 G_F}{\sqrt{2} \pi} V_{ts}^* V_{tb} \left[ \sum_{i=1}^{6} C_i(\mu) \mathcal{O}_i(\mu) + C_7 \frac{e}{16 \pi^2} [\overline{s} \sigma_{\mu \nu}(m_s P_L  + m_b P_R)b]F^{\mu \nu} \right. \nonumber \\ 
& & \left.+ C_9 \frac{\alpha_{em}}{4 \pi}(\overline{s} \gamma^{\mu} P_L b)(\overline{\ell} \gamma_{\mu} \ell) + C_{10} \frac{\alpha_{em}}{4 \pi} (\overline{s} \gamma^{\mu} P_L b)(\overline{\ell} \gamma_{\mu} \gamma_5 \ell) \right],
\end{eqnarray} 
where $G_F$ is the Fermi constant, $V_{ts}$ and $V_{tb}$ are the Cabibbo-Kobayashi-Maskawa (CKM) matrix elements and $P_{L,R} = (1 \mp \gamma^{5})/2$ are the projection operators. The effect of the operators $\mathcal{O}_i,\,i=1-6,8 $ can be embedded in the redefined effective Wilson coefficients (WCs) as $C_7(\mu)\rightarrow C^{\rm eff}_7(\mu,q^2)$ and $C_9(\mu)\rightarrow C^{\rm eff}_9(\mu,q^2)$. 

We now add following NP contributions to the SM effective Hamiltonian,
\begin{eqnarray}
\mathcal{H}^{\rm NP}_{\rm VA} &=& -\frac{\alpha_{\rm em} G_F}{\sqrt{2} \pi} V_{ts}^* V_{tb} \left[ C^{\rm NP,\,e}_9 \, (\overline{s} \gamma^{\mu} P_L b)\, (\overline{e} \gamma_{\mu} e) + C^{\rm NP,\, e}_{10} \, (\overline{s} \gamma^{\mu} P_L b)\, (\overline{e} \gamma_{\mu} \gamma_5 e) \right. \nonumber \\
& & +\left. C^{\prime, \, \rm e}_9 \, (\overline{s} \gamma^{\mu} P_R b)\, (\overline{e} \gamma_{\mu} e) + C^{\prime, \, \rm e}_{10} \, (\overline{s} \gamma^{\mu} P_R b)
\, (\overline{e} \gamma_{\mu} \gamma_5 e)\right], \label{heff-va}\\
\mathcal{H}^{\rm NP}_{\rm SP} &=& -\frac{\alpha_{\rm em}G_F}{\sqrt{2}\pi}V^*_{ts}V_{tb}
\left[ C^{\rm e}_{SS} \, (\overline{s}  b)(\overline{e}  e) + C^{\rm e}_{SP} \, (\overline{s}  b)(\overline{e} \gamma_5 e) \right. \nonumber \\
& & + \left. C^{\rm e}_{PS} \, (\overline{s} \gamma_5 b)\,(\overline{e}  e) + C^{\rm e}_{PP} \, (\overline{s} \gamma_5 b)\,(\overline{e} \gamma_5 e)\right],\label{heff-sp}\\
\mathcal{H}^{\rm NP}_{\rm T} &=& -\frac{\alpha_{\rm em}G_F}{\sqrt{2}\pi}V^*_{ts}V_{tb}
\left[ C^{\rm e}_T\, (\overline{s}  \sigma^{\mu \nu} b)\,(\overline{e} \sigma_{\mu \nu}  e) + C^{\rm e}_{T5} \, (\overline{s} \sigma^{\mu \nu}  b)\, (\overline{e} \sigma_{\mu \nu}  \gamma_5 e) \right] \label{heff-t},
\end{eqnarray} 
where $C^{\rm NP,\, e}_{9,10}$, $C^{\prime, \, \rm e}_{9,10}$,  and $C^{\rm e}_{SS, SP, PS, PP, T, T5}$ are the NP WCs. 

Using simple symmetry arguments, we can argue that S/P operators can not provide a solution to 
$R_K/R_{K^*}$ discrepancy. The operators containing the quark bilinear $\bar{s}b$
    can lead to $B \to K e^+ e^-$ transition but not
    $B \to K^* e^+ e^-$ transition. Such operators can
account for $R_K$ but not $R_{K^*}$. On the other hand, the operators containing the quark pseudoscalar
bilinear $\overline{s} \gamma_5 b$ can not lead to $B \to K e^+ e^-$. These operators can not account for $R_K$. 
In addition, the contribution of these operators to $B_s \to e^+ e^-$ is not subject to helicity suppression. Hence, the
coefficients $C^{\rm e}_{PS}$ and $C^{\rm e}_{PP}$ are constrained to be very small. The current upper 
limit on $\mathcal{B} (B_s\to e^+e^-) <9.4\times 10^{-9}$ at $90\%$ C.L., leads to the condition
\begin{equation}
|C^{\rm e}_{PS}|^2 + |C^{\rm e}_{PP}|^2 \lesssim 0.01,
\label{bsee}
\end{equation}
whereas one needs 
\begin{equation}
120 \lesssim |C^{\rm e}_{PS}|^2 + |C^{\rm e}_{PP}|^2 \lesssim 345,\quad \, 9 \lesssim |C^{\rm e}_{PS}|^2 + |C^{\rm e}_{PP}|^2 \lesssim 29,
\label{psops}
\end{equation}
to satisfy the experimental constraint on $R^{\rm low}_{K^*}$ and $R^{\rm central}_{K^*}$ respectively.
Therefore, we will not consider S/P operators in our fit procedure.

The NP Hamiltonian can potentially impact  observables in the decays induced by the quark level transition $b\to se^+e^-$. To obtain the values of NP WCs, we perform a fit to the current data in $b\to se^+e^-$ sector. We consider following fifteen observables in our fit:
\begin{itemize}
\item Measured values of $R_K$ in $1.1\leq q^2\leq 6.0$ GeV$^2$ bin ~\cite{Aaij:2019wad}  and $R_{K^*}$ in both $0.045 < q^2 < 1.1$  GeV$^2$ and $1.1 < q^2 < 6.0 $ GeV$^2$ bins  by the LHCb collaboration~\cite{Aaij:2017vbb},
\item Measured values of $R_{K^*}$ by the Belle collaboration in $0.045 < q^2 < 1.1$  GeV$^2$, $1.1 < q^2 < 6.0 $ GeV$^2$ and $15.0 < q^2 < 19.0 $ GeV$^2$ bins for both $B^0$ and $B^+$ decay modes~\cite{Abdesselam:2019wac},
\item The upper limit of $\mathcal{B}(B_s\to e^+e^-)< 9.4\times 10^{-9}$ at $90\%$ C.L. by the LHCb collaboration~\citep{Aaij:2020nol},
\item The differential branching fraction of $B\to K^*e^+e^-$, $(3.1^{+0.9}_{-0.8} \pm 0.2)\times 10^{-7}$, in $0.001< q^2 <1.0$ GeV$^2$ bin  by the LHCb collaboration~\cite{Aaij:2013hha},
\item The measured value of $K^*$ longitudinal polarization fraction $F_L$, $0.16\pm 0.06\pm 0.03$,  in $0.002< q^2 <1.12$ GeV$^2$ bin by the LHCb collaboration~\cite{Aaij:2015dea}\footnote{We do  not include the recent measurement of $F_L$ by the LHCb collaboration ~\cite{Aaij:2020umj} in the $q^2$ bin (0.0008 - 0.257) GeV$^2$.},
\item Measured values of the branching ratios of $B\to X_se^+e^-$ by the BaBar collaboration in both $1.0<q^2<6.0$ GeV$^2$ and $14.2<q^2 <25.0$ GeV$^2$ bins which are $\left(1.93^{+0.47+0.21}_{-0.45-0.16}\pm 0.18\right)\times 10^{-6}$ and $\left(0.56^{+0.19+0.03}_{-0.18-0.03}\right)\times 10^{-6}$, respectively~\cite{Lees:2013nxa},
\item Measured values of $P^{\prime}_4$ in $B\to K^* e^+e ^-$ decay by the Belle collaboration in $1.0< q^2 <6.0$ GeV$^2$ and $14.18< q^2 <19.0$ GeV$^2$  bins which are $-0.72^{+0.40}_{-0.39}\pm 0.06$ and $-0.15^{+0.41}_{-0.40}\pm 0.04$, respectively~\cite{Wehle:2016yoi},
\item Measured values of $P^{\prime}_5$ in $B\to K^* e^+e ^-$ decay by the Belle collaboration in $1.0< q^2 <6.0$ GeV$^2$ and $14.18< q^2 <19.0$ GeV$^2$  bins which are $-0.22^{+0.39}_{-0.41}\pm 0.03$ and $-0.91^{+0.36}_{-0.30}\pm 0.03$, respectively ~\cite{Wehle:2016yoi}.
\end{itemize}
We define the $\chi^2$ function as
\begin{equation}
\chi^2 (C_i) = \sum_{\rm all\, obs.} \frac{\left(O^{\rm th}(C_i) - O^{\rm exp}\right)^2}{\sigma^2_{\rm exp}+ \sigma^2_{\rm th}}.
\end{equation}
Here $O^{\rm th}(C_i)$ are the theoretical predictions of the observables taken into fit which depend on the NP WCs and $O^{\rm exp}$ are the measured central values of the corresponding observables. The $\sigma_{\rm exp}$ and $\sigma_{\rm th}$ are the experimental and theoretical uncertainties, respectively. The experimental errors in all observables dominate over the theoretical errors. In case of the asymmetric errors, we use the larger error in our analysis. 
The prediction of $O^{\rm th}(C_i)$ is obtained  using {\tt Flavio} package~\cite{Straub:2018kue} which uses the most precise form factor predictions obtained in the light cone sum rule (LCSR)~\cite{Straub:2015ica, Gubernari:2018wyi} approach. The non-factorisable corrections are incorporated following the parameterization used in Ref.~\cite{Straub:2015ica,Straub:2018kue}. These are also compatible with the calculations in Ref.~\cite{Khodjamirian:2010vf,Gubernari:2020eft}.
 
We obtain the values of NP WCs by minimizing the $\chi^2$ using CERN minimization code {\tt Minuit}~\cite{James:1975dr,James:1994vla}. We perform the minimization in two ways: (a) one NP operator at a time and (b) two NP operators at a time. 
Since we do the fit with fifteen data points, it is expected that an NP scenario with a value of $\chi^2_{\rm min} \approx 15$
provides a good fit to the data. We also define pull $=\sqrt{\Delta \chi^2}$ where $\Delta \chi^2 = (\chi^2_{\rm SM} - \chi^2_{\rm min})$. Since $\chi^2_{\rm SM}
\approx 27$, any scenario with pull $\gtrsim 3.0$ can be considered to be a viable solution. 
 In the next section, we present our fit results and discuss them in details.

\section{Vector/axial-vector new physics}
\label{VA}

There are four cases for one operator fit and six cases for two operators fit. For all of these cases, we list the best fit values of  WCs in Table~\ref{tab1} along with their $\chi^2_{\rm min}$ values. We also calculate the corresponding values of pull which determine the degree of improvement over the SM.
\begin{table}[ht!]
	\begin{center}
		\begin{tabular}{|c|c|c|c|}
			\hline \hline
Wilson Coefficient(s) &	Best fit value(s) &  $\chi^2_{\rm min}$ & pull \\ \hline
$C_i = 0$ (SM) & $-$ & 27.42  &\\ \hline
\multicolumn{4}{|c|}{1D Scenarios} \\ \hline
$C^{\rm NP ,e}_9$& $0.91 \pm 0.28 $ & 15.21 & 3.5\\ \hline
$C^{\rm NP ,e}_{10}$& $-0.86 \pm 0.25 $ & 12.60   & 3.8\\ \hline
$C^{\prime ,e }_9$& $0.24\pm 0.24 $ & 26.40 & 1.0\\ \hline
$C^{\prime ,e}_{10}$& $-0.17 \pm 0.21 $ & 26.70  & 0.8\\ \hline
\multicolumn{4}{|c|}{2D Scenarios} \\ \hline
$(C^{\rm NP ,e}_9, C^{\rm NP ,e}_{10})$ & $(-1.03, -1.42)$ & 11.57 & 3.9\\ \hline

$(C^{\rm NP ,e}_9,C^{\prime ,e}_{9})$&$(-3.61, -4.76)$ & 17.65 & 3.1\\ 
&$(-3.52, 4.29)$ & 15.71 & 3.4\\ 
&$(1.21, -0.54)$ & 12.83 & 3.8\\ \hline
$(C^{\rm NP ,e}_9,C^{\prime ,e}_{10})$&$(1.21, 0.69)$ & 12.39 & 3.9 \\	\hline	
	
	$( C^{\prime ,e }_9, C^{\rm NP ,e}_{10})$& $(-0.50, -1.03)$ & 11.30 & 4.0\\	 \hline	
	
$(C^{\prime ,e}_9,C^{\prime ,e}_{10})$&$(2.05, 2.33)$ & 10.41 & 4.1\\ 
&$(-2.63, -1.86)$ & 12.71 & 3.8\\ \hline

$(C^{\rm NP ,e}_{10},C^{\prime ,e}_{10})$&$(3.64, 5.33)$ & 18.50 & 3.0\\		
&$(-1.04, 0.38)$ & 11.14 & 4.0\\	
&$(4.56,-5.24)$ & 16.58 & 3.3\\ \hline \hline	
		\end{tabular}
	\caption{The best fit values of NP WCs in $b\rightarrow se^+e^-$ transition for 1D and 2D scenarios.  The value of $\chi^2_{\rm SM}$ is  27.42. }
\label{tab1}
\end{center} 
\end{table} 

From Table~\ref{tab1}, we find that the $C_9^{\rm NP, e}$ and $C_{10}^{\rm NP ,e}$ scenarios provide a  good fit to the $b\to s e^+e^-$ data. However, the other two 1D scenarios, $C_9^{\prime ,e}$ and $C_{10}^{\prime ,e}$, fail to provide any improvement over the SM. Therefore, we reject them on the basis of $\Delta \chi^2$ or pull. In the case of 2D framework, all six combinations improve the global fit  as compared to the SM.

\begin{figure*}[ht] 
\centering
\includegraphics[width=75mm]{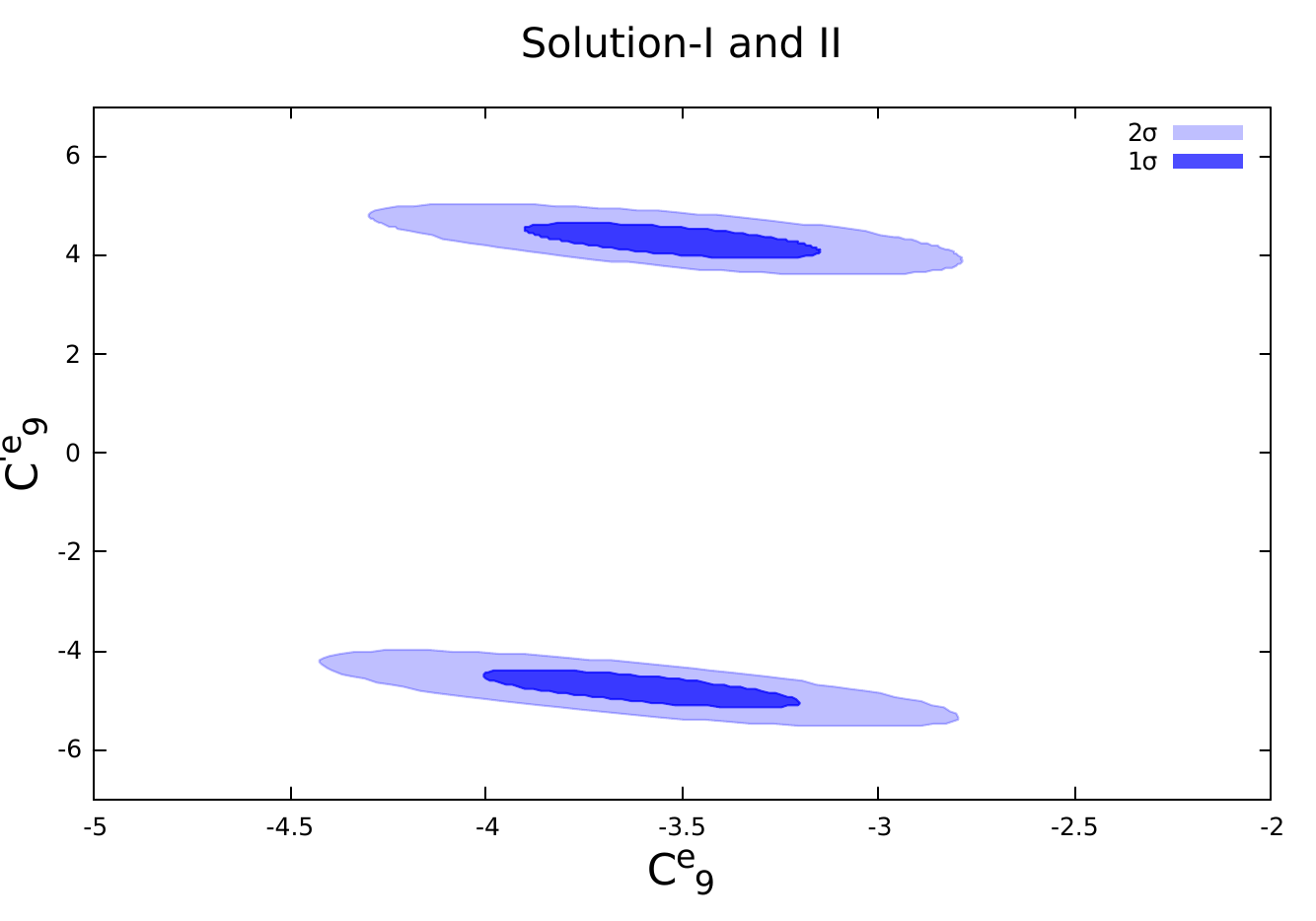}
\hspace{0.65cm}
\includegraphics[width=75mm]{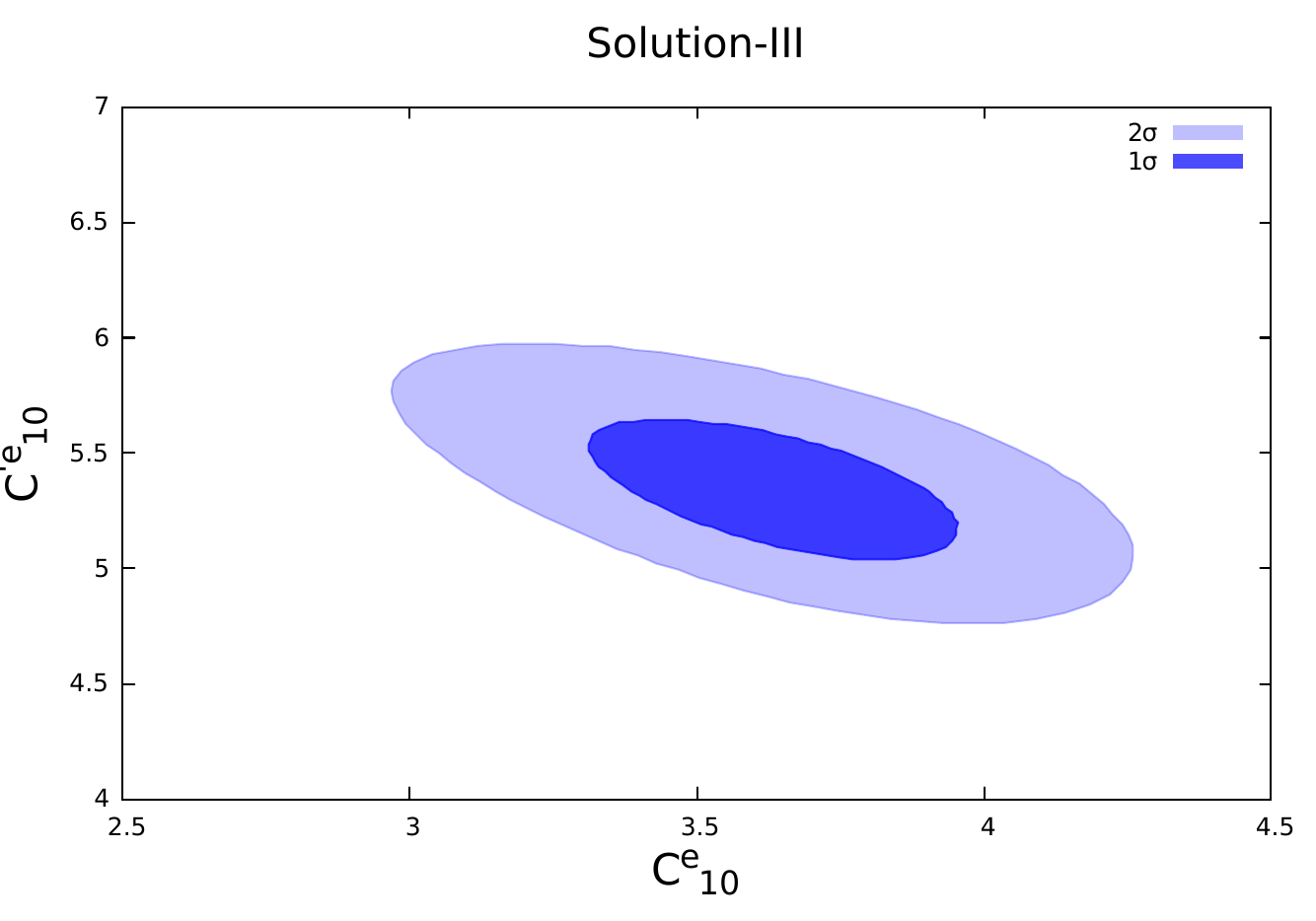}\\
\caption{The allowed $1\sigma$ and $2\sigma$ ellipses for the three 2D solutions listed in Table~\ref{tab3}.}
\label{fig1}
\end{figure*}

We now impose the stringent condition that a NP solution must predict the values of $R_K$, $R^{\rm low}_{K^*}$ and 
$R^{\rm central}_{K^*}$ to be within 1$\sigma$ of their measured values. 
In order to identify solutions satisfying this condition, we calculate the predictions of $R_K/R_{K^*}$ for all good fit scenarios. The predicted values of these quantities are listed in Table~\ref{tab2} from which we observe that the 1D  scenario $C_9^{\rm NP}$ could not accommodate both the $R^{\rm low}_{K^*}$ and $R^{\rm central}_{K^*}$ within $1\sigma$ whereas most of the other solutions fail to explain the $1\sigma$ range of $R^{\rm low}_{K^*}$ only. There are only three 2D solutions whose predictions for $R_K$, $R^{\rm low}_{K^*}$ and $R^{\rm central}_{K^*}$ are within $1\sigma$ of their measurements. We call these scenarios as allowed NP solutions and list them in Table~\ref{tab3}. The $1\sigma$ and $2\sigma$ allowed regions for these three allowed solutions are shown in Fig~\ref{fig1}.
\begin{table}[ht!]
	\begin{center}
		\begin{tabular}{|c|c|c|c|c|c|}
			\hline \hline
Wilson Coefficient(s) &	Best fit value(s) &pull & $R_K$ & $R^{\rm low}_{K^*}$ & $R^{\rm central}_{K^*}$ \\ \hline \hline	
\multicolumn{3}{|c|}{Expt. $1\sigma$ range}& $[0.784,0.908]$ & $[0.547,0.773]$ &  $[0.563,0.807]$ \\
 \hline\hline
 \multicolumn{6}{|c|}{1D Scenarios} \\ \hline
$C^{\rm NP ,e}_9$& $0.91 \pm 0.28 $ &  3.5 & $0.806 \pm 0.001$ & $0.883\pm 0.008$ & $0.832\pm 0.009$\\ \hline
$C^{\rm NP ,e}_{10}$& $-0.86 \pm 0.25 $ & 3.8 & $0.805\pm 0.005$ & $0.855\pm 0.007$ & $0.778\pm 0.012$ \\ \hline
 \multicolumn{6}{|c|}{2D Scenarios} \\ \hline
$(C^{\rm NP ,e}_9, C^{\rm NP ,e}_{10})$ & $(-1.03, -1.42)$ &3.9  &     $0.825\pm 0.011$ & $0.832\pm 0.007$ & $0.745\pm 0.026$    \\ \hline

$(C^{\rm NP ,e}_9,C^{\prime ,e}_{9})$&$(-3.61, -4.76)$ &   3.1   & $0.867\pm 0.050$ & $0.757\pm 0.007$ & $0.625\pm 0.024$  \\ 
&$(-3.52, 4.29)$ & 3.4 & $0.832\pm 0.001$ & $0.798\pm 0.028$ & $0.707\pm 0.090$   \\ 
&$(1.21, -0.54)$ &  3.8  & $0.853\pm 0.001$& $0.825\pm 0.018$ & $0.701\pm 0.012$   \\ \hline
$(C^{\rm NP ,e}_9,C^{\prime ,e}_{10})$&$(1.21, 0.69)$ & 3.9    & $0.855\pm 0.004$& $0.819\pm 0.016$ &  $0.691\pm 0.011$    \\	\hline	
	
	$( C^{\prime ,e}_9, C^{\rm NP ,e}_{10})$& $(-0.50, -1.03)$ &  4.0   & $0.844\pm 0.007$ & $0.812\pm 0.012$ & $0.690\pm 0.009$   \\	 \hline

$(C^{\prime ,e}_9,C^{\prime ,e}_{10})$&$(2.05, 2.33)$ &   4.1   & $0.845\pm 0.010$ & $0.808\pm 0.014$ & $0.683\pm 0.029$  \\ 
&$(-2.63, -1.86)$ &  3.8   & $0.856\pm 0.020$& $0.808\pm 0.015$ & $0.684\pm 0.010$  \\ \hline

$(C^{\rm NP ,e}_{10},C^{\prime ,e}_{10})$&$(3.64, 5.33)$ & 3.0 & $0.860\pm 0.015$ & $0.788\pm 0.014$ & $0.645\pm 0.015$   \\		
&$(-1.04, 0.38)$ &   4.0   &$0.846\pm 0.004$ & $0.809\pm 0.013$ & $0.686\pm 0.014$  \\	
&$(4.56,-5.24)$ & 3.3   &  $0.842\pm 0.004$& $0.809\pm 0.015$ & $0.685\pm 0.019$    \\ \hline \hline
		\end{tabular}
	\caption{The predictions of $R_K$, $R^{\rm low}_{K^*}$ and $R^{\rm central}_{K^*}$ for the good fit scenarios obtained in Table.~\ref{tab1}.}
\label{tab2}
\end{center} 
\end{table}

\begin{table}
	\begin{center}
		\begin{tabular}{|c|c|c|c|c|c|c|}
			\hline \hline
Solution & Wilson Coefficient(s) &	Best fit value(s) & pull & $R_K$ & $R^{\rm low}_{K^*}$ & $R^{\rm central}_{K^*}$ \\ \hline
\hline	
\multicolumn{4}{|c|}{Expt. $1\sigma$ range} & $[0.784,0.908]$ & $[0.547,0.773]$ &  $[0.563,0.807]$ \\ \hline \hline
\multicolumn{7}{|c|}{2D Scenarios} \\ \hline
I& $(C^{\rm NP ,e}_9,C^{\prime ,e}_{9})$&$(-3.61, -4.76)$ & 3.1   & $0.867\pm 0.050$ & $0.757\pm 0.007$ & $0.625\pm 0.024$  \\ 
II& &$(-3.52, 4.29)$ & 3.4  & $0.832\pm 0.001$ & $0.798\pm 0.028$ & $0.707\pm 0.090$   \\ \hline
			
III & $(C^{\rm NP ,e}_{10},C^{\prime ,e}_{10})$&$(3.64, 5.33)$ & 3.0 & $0.860\pm 0.015$ & $0.788\pm 0.014$ & $0.645\pm 0.015$   \\		 \hline \hline	
		\end{tabular}
	\caption{Here we list only those NP WCs which generate $R_K$ and $R_{K^*}$ within $1\sigma$ range of their experimental values.}
\label{tab3}
\end{center} 
\end{table}

The EFT analysis can serve as a guideline for constructing NP models. A detailed analysis  of possible models which can generate the favoured Lorentz structure obtained above is beyond the scope of this work. Here we briefly discuss some of the  simple models
which can generate these scenarios at the tree level. The pattern of NP obtained through the above model independent analysis can be realised in those NP models where $b \to s \mu^+ \mu^-$  transition remains SM like. Hence models based on $L_{\mu}-L_{\tau}$ gauge symmetry \cite{Heeck:2011wj,Altmannshofer:2014cfa,Crivellin:2015mga,Altmannshofer:2015mqa,Crivellin:2016ejn} as well as partial compositeness \cite{Niehoff:2015bfa} would not generate the allowed Lorentz structures as these models naturally generate NP effects in the muon sector, while keeping $b \to s e^+ e^-$ SM like.   The allowed EFT scenarios can be generated in a  $Z'$ model with coupling only to electrons and avoiding LEP constraints. For e.g., a light $Z'$ ($M_{Z'}\sim $ 25 MeV) with a $q^2$ dependent $b-s$ coupling  that couples to the electron but not to the muons can induce the favored operators \cite{Datta:2017ezo}. Another alternative would be a class of scalar or vector leptoquark models  with  coupling only to electrons along  with 
flavor-changing quark couplings, see for e.g., \cite{Hiller:2014yaa,DAmico:2017mtc, Dorsner:2016wpm}.

After identifying the allowed solutions, we find out the set of observables which can  discriminate  between them. In the next 
subsection, we investigate discriminating capabilities of the standard angular observables in $B \to K^* e^+ e^-$ decay.

\subsection{Discriminating  V/A solutions  }
\label{unique}

The differential distribution of the four-body decay $B\to K^*(\to K\pi)e^+e^-$ can be parametrized as the function of one kinematic ($q^2$) and three angular variables $\overrightarrow{\Omega} = (\cos\theta_K, \cos\theta_e, \phi)$. The kinematic variable is $q^2 = (p_B-p_{K^*})^2$, where $p_B$ and $p_{K^*}$ are respective four-momenta of $B$ and $K^*$ mesons. The angular variables are defined in the $K^*$ rest frame. They are (a) $\theta_{K}$ the angle between 
$B$ and $K$ mesons where $K$ meson comes from $K^*$ decay, (b) $\theta_{e}$ the angle between momenta of $e^-$ and $B$ meson and (c) $\phi$ the angle between $K^*$ decay plane and the plane defined by the $e^+-e^-$ momenta. The CP averaged angular distribution of the $B\to K^*(\to K\pi)e^+e^-$ decay can be written as~\cite{Aaij:2015oid}
\begin{eqnarray}
\frac{1}{d(\Gamma +\bar{\Gamma})/dq^2}\frac{d^4(\Gamma +\bar{\Gamma})}{dq^2d\overrightarrow{\Omega}} &= & \frac{9}{32\pi}\left[\frac{3}{4}\left(1-F_L\right)\sin^2\theta_K + F_L \cos^2\theta_K + \frac{1}{4}(1-F_L) \sin^2\theta_K \cos 2\theta_e \right. \nonumber\\
& &-F_L \cos^2\theta_K \cos 2\theta_e +S_3 \sin^2\theta_K \sin^2 \theta_e \cos 2\phi \nonumber \\
& & + S_4 \sin 2\theta_K \sin 2\theta_e \cos \phi + S_5 \sin 2\theta_K \sin \theta_e \cos \phi \nonumber \\
& & +\frac{4}{3} A_{FB} \sin^2 \theta_K \cos\theta_e + S_7 \sin2\theta_K \sin\theta_e \sin\phi \nonumber \\
& & + S_8 \sin 2\theta_K \sin 2\theta_e \sin \phi +S_9 \sin^2 \theta_K \sin^2 \theta_e \sin 2\phi \Big].
\end{eqnarray}
Following the notations of ref.~\cite{Altmannshofer:2008dz}, the $q^2$ dependent $CP$ averaged angular observables $S_i$ can be defined as
\begin{equation}
S_i(q^2) = \frac{I_i(q^2)+ \bar{I}_i(q^2)}{d(\Gamma +\bar{\Gamma})/dq^2}.
\end{equation}
The detailed expressions of angular coefficients $I_i$ can also be found in ref.~\cite{Altmannshofer:2008dz}.

The longitudinal polarization fraction of $K^*$, $F_L$, depends on the distribution of the events in the angle $\theta_K$ (after integrating over $\theta_e$ and $\phi$) and the forward-backward asymmetry, $A_{FB}$, is defined in terms of $\theta_e$ (after integrating over $\theta_K$ and $\phi$). We can write these two quantities in terms of $S^{(a)}_i$ as follows~\cite{Altmannshofer:2008dz}
\begin{equation}
A_{FB} = \frac{3}{8}\left(2S^s_6+S^c_6\right), \quad F_L = -S^c_2.
\end{equation}
In addition to the $S_i$ observables, one can also investigate the NP effects on a set of optimized observables $P_i$. In fact, the observables $P_i$ are theoretically cleaner in comparison to the  form factors dependent observables $S_i$. These two sets of observables are related to each other \cite{DescotesGenon:2012zf}. However, there are several notations used in the literature. The definition of the $P_i$ observables used in this work follows the LHCb convention~\cite{Aaij:2015oid}
\begin{equation}
P_1 = \frac{2S_3}{1-F_L}, \quad P_2 = \frac{2}{3}\frac{A_{FB}}{(1-F_L)}, \quad P_3 = \frac{-S_9}{1-F_L}, \nonumber
\end{equation}
\begin{equation}
P^{\prime}_4 = \frac{S_4}{\sqrt{F_L(1-F_L)}}, \quad P^{\prime}_5 = \frac{S_5}{\sqrt{F_L(1-F_L)}},\quad P^{\prime}_6 = \frac{S_7}{\sqrt{F_L(1-F_L)}}, \quad P^{\prime}_8 = \frac{S_8}{\sqrt{F_L(1-F_L)}}.
\label{Pi}
\end{equation}
The measurements of $P^{\prime}_4$ and $P^{\prime}_5$  observables by the Belle collaboration \cite{Wehle:2016yoi} used in our fits also follow the LHCb notation. A complete relations between the LHCb definitions and the notations used in different papers can be found in ref.~\cite{Gratrex:2015hna}.

\begin{table}[ht!]
\begin{tabular}{c|c|c|c|c|c}
\hline
Observable & $q^2$ bin & SM & S-I & S-II & S-III  \\
  \hline
 $A_{FB}$  & $[1.1, 6]$ & $0.008\pm 0.031$ & $-0.146\pm 0.026$ & $-0.161\pm 0.027$ & $-0.016\pm 0.011$ \\
 & $[15,19]$ & $0.368\pm 0.018$ & $-0.005\pm 0.003$ & $0.002\pm 0.005$ & $0.026\pm 0.004$\\
 \hline
  $F_L$  & $[1.1, 6]$ & $0.764\pm 0.043$ & $0.630\pm 0.056$ & $0.599\pm 0.055$ & $0.765\pm 0.042$ \\
 & $[15,19]$ & $0.341\pm 0.020$ & $0.338\pm 0.022$ & $0.325\pm 0.020$ & $0.349\pm 0.020$ \\
 \hline
\end{tabular}
\caption{Average values of $B \to K^* e^+ e^-$ angular observables  $A_{FB}$ and $F_L$ in SM as well as for the allowed NP V/A solutions listed in Table.~\ref{tab3}.}
\label{tab-FLAFB}
\end{table}

\begin{figure}[ht!]
\centering
\begin{tabular}{cc}
\includegraphics[width=75mm]{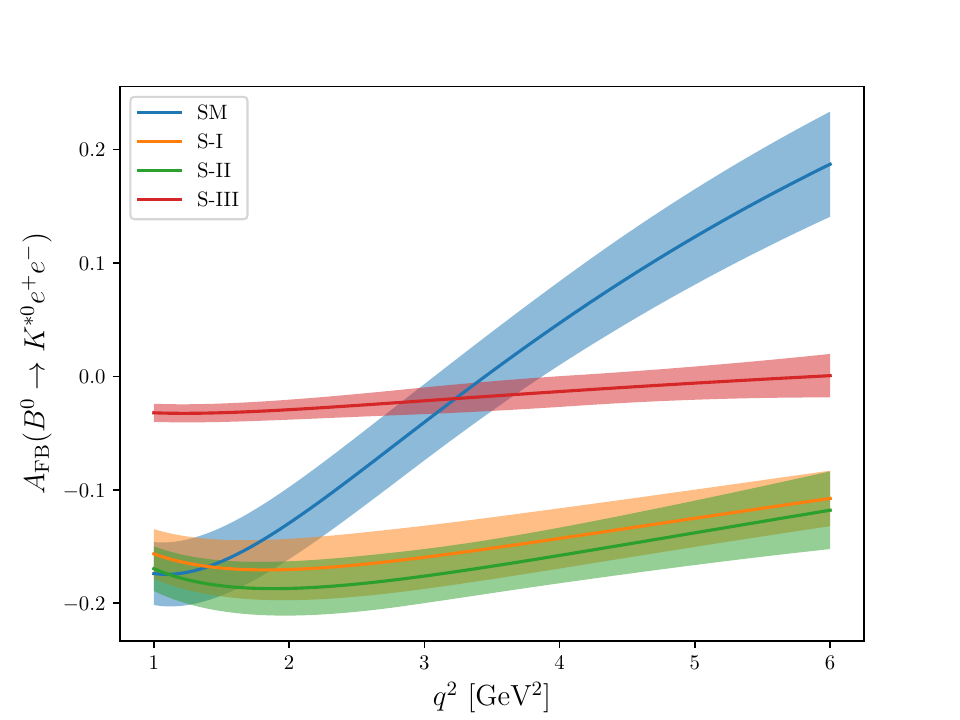} &
\includegraphics[width=75mm]{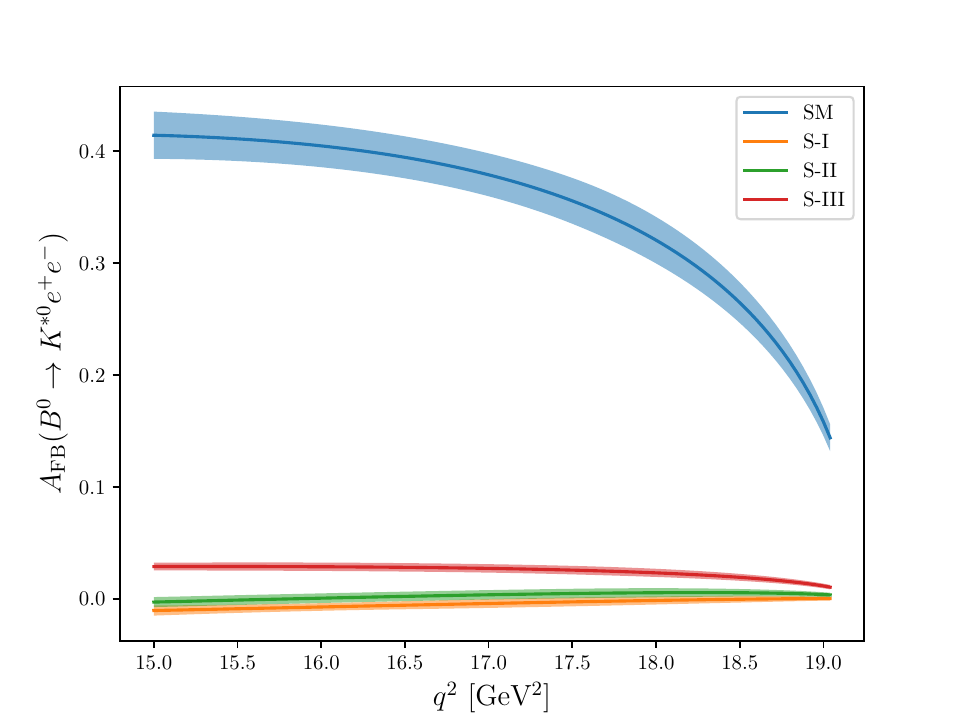}\\
\includegraphics[width=75mm]{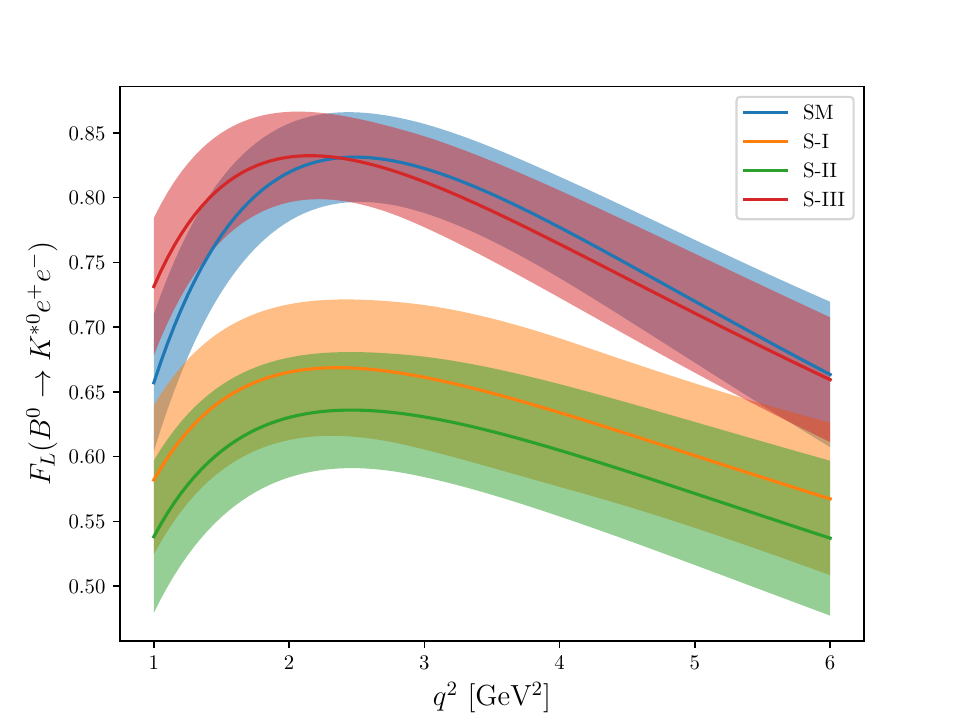} &
 \includegraphics[width=75mm]{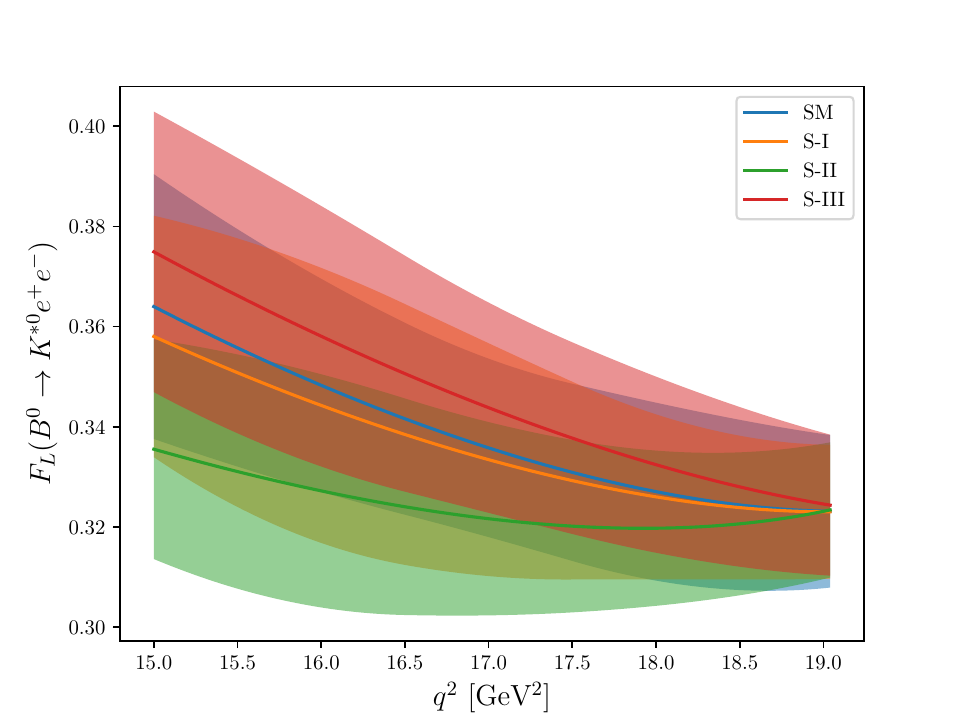}\\
 \end{tabular}
\caption{Plots of $A_{FB}$ and $F_L$ as a function of $q^2$ for the SM and the three NP V/A solutions. The left and right panels correspond to the low ($[1.1,6.0]$ GeV$^2$) and high ($[15,19]$ GeV$^2$) $q^2$ bins, respectively.}
\label{fig2}
\end{figure}

We calculate $A_{FB}$, $F_L$ along with  optimized observables $P_{1,2,3}$ and $P^{\prime}_{4,5,6,8}$ for the SM and the three allowed NP solutions in $q^2\subset [1.1,6.0]$ and $[15,19]$ GeV$^2$ bins. The average values of  $A_{FB}$ and $F_L$ are listed in Table~\ref{tab-FLAFB} and the $q^2$ plots are shown in Fig.~\ref{fig2}. From the predictions, we observe the following features:
\begin{itemize}
\item In low $q^2$ region, the SM prediction of $A_{FB}(q^2)$ has a zero crossing at $\sim 3.5$ GeV$^2$. For the NP solutions, the predictions are negative throughout the low $q^2$ range. However, the  $A_{FB}(q^2)$ curve is almost the same for S-I and S-II whereas for S-III, it  is markedly different. Therefore an accurate measurement of $q^2$ distribution of $A_{FB}$ can discriminate between S-III and the remaining two NP solutions.
\item
In high $q^2$ region, the SM prediction of $A_{FB}$ is $0.368\pm 0.018$ whereas the predictions for the three solutions 
are almost zero. If $A_{FB}$ in high $q^2$ region is measured to be small, it provides additional confirmation for the existence 
of NP, which is indicated by the reduced values of $R_K$ and $R_{K^*}$.
All the three NP solutions induce a large deviation in  $A_{FB}$, but the discriminating capability of $A_{FB}$ is extremely limited. 
\item The  S-I and S-II  scenarios can marginally suppress the value of  $F_L$ in low $q^2$ region compared to the SM whereas for  S-III, the predicted value is consistent with the SM.  In high $q^2$ region, $F_L$ for all three scenarios are close to the SM value. Hence $F_L$ cannot discriminate between the allowed V/A solutions.
\end{itemize}

\begin{figure}[ht]
\centering
\begin{tabular}{cc}
\includegraphics[width=75mm]{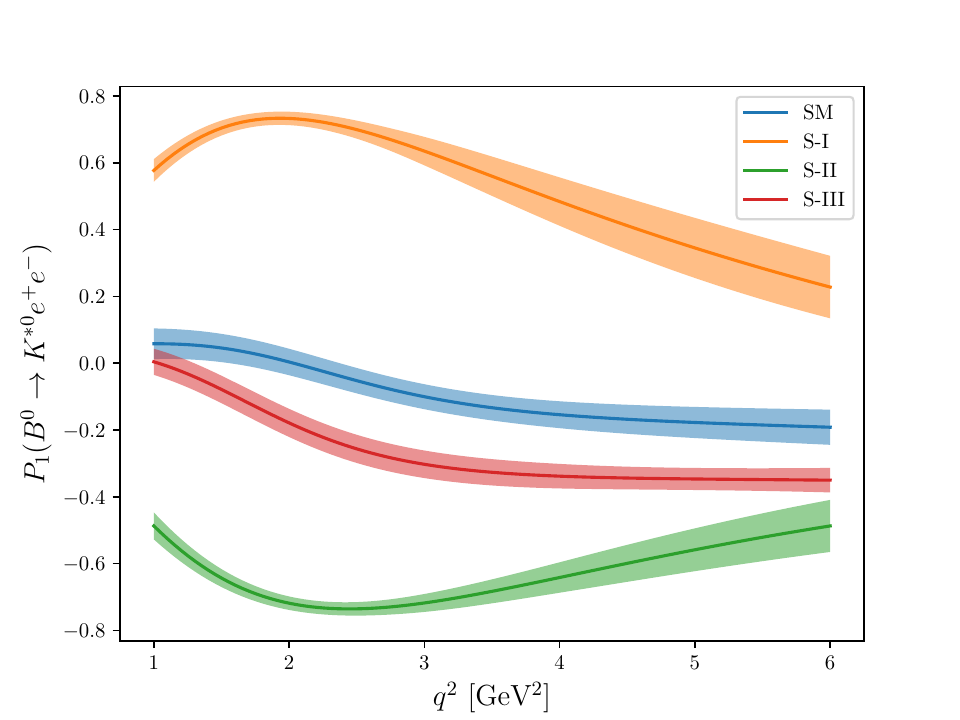} &
 \includegraphics[width=75mm]{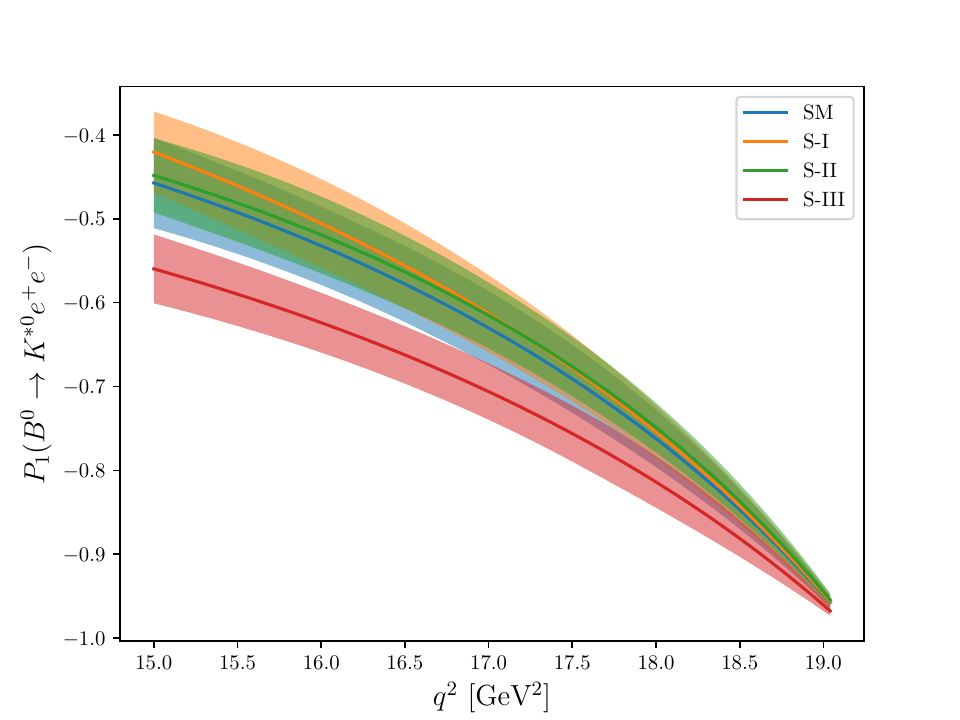}\\
\includegraphics[width=75mm]{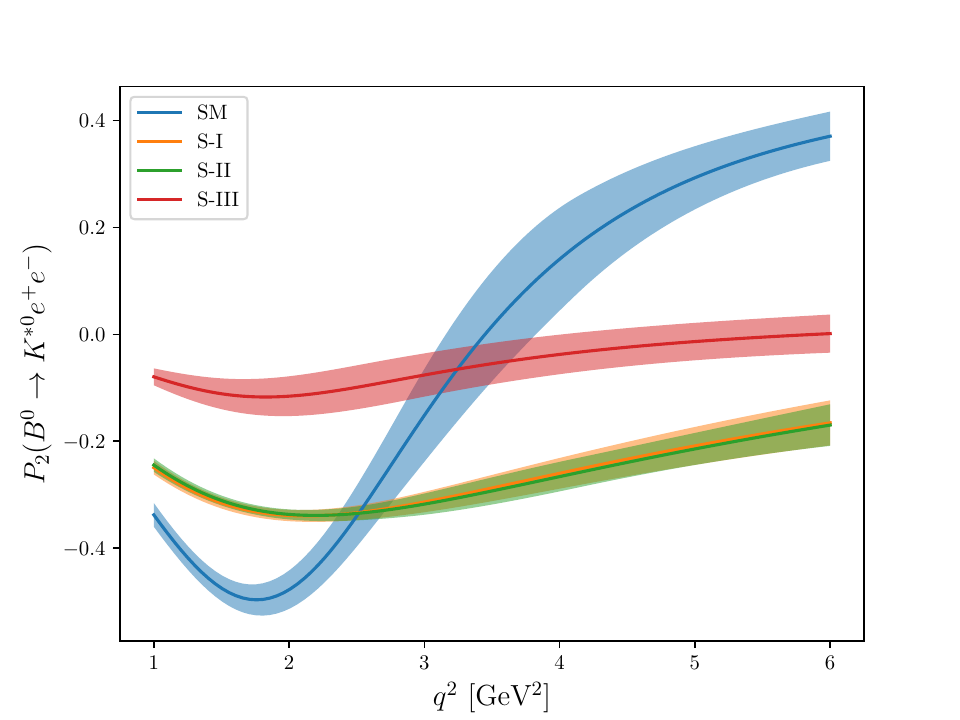} &
 \includegraphics[width=75mm]{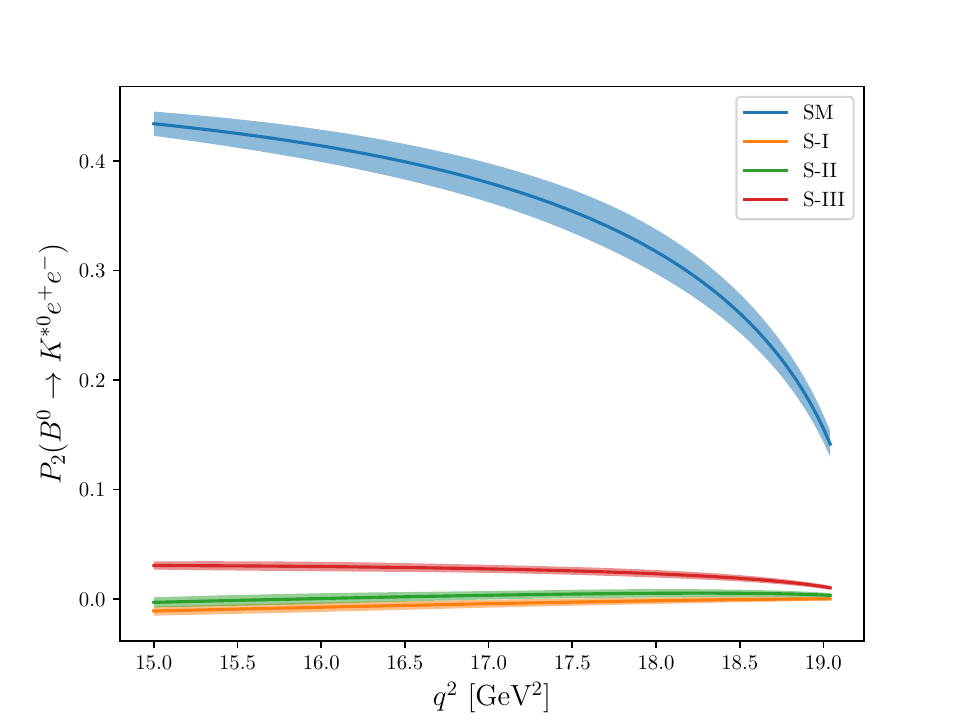}\\
\includegraphics[width=75mm]{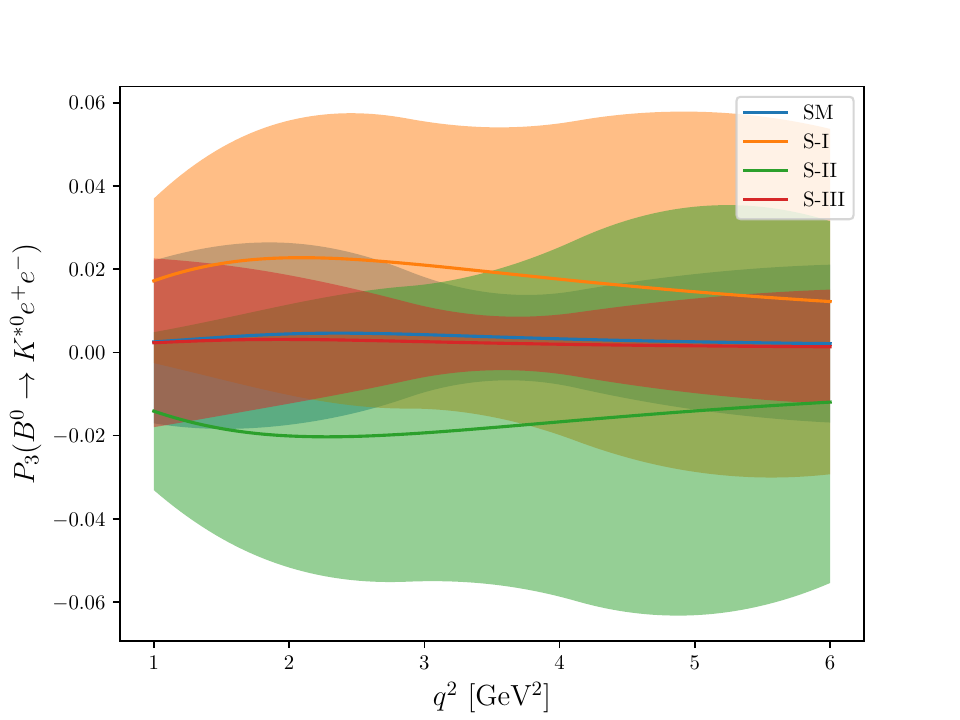} &
 \includegraphics[width=75mm]{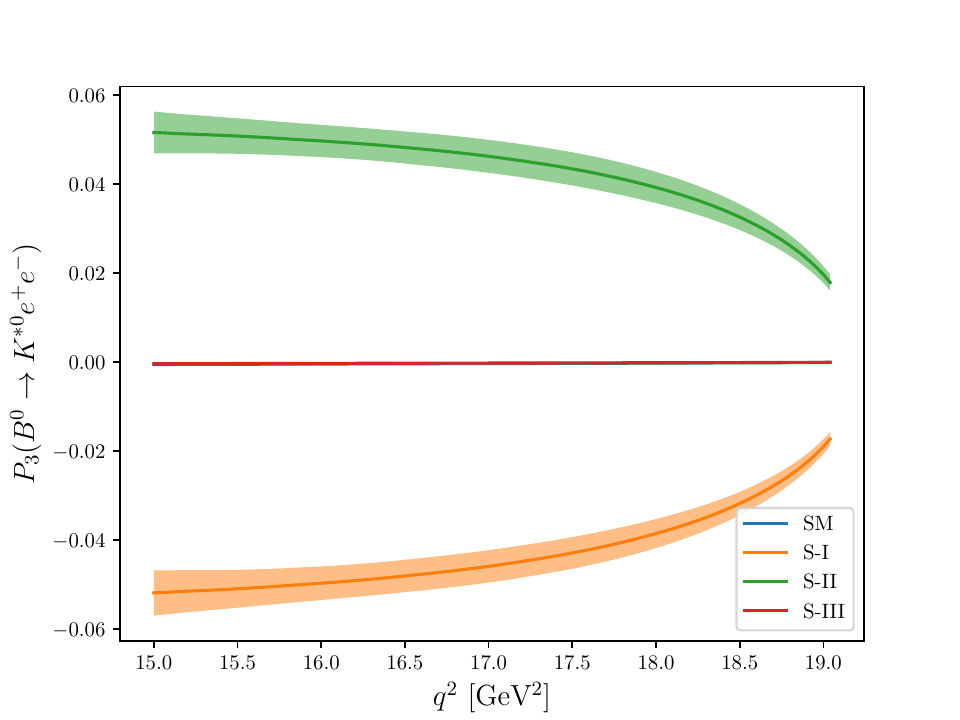}\\
 \end{tabular}
\caption{Plots of $P_{1,2,3}(q^2)$ as a function of $q^2$ for the SM and three NP scenarios. The left and right panels correspond to the low ($[1.1,6.0]$ GeV$^2$) and high ($[15,19]$ GeV$^2$) $q^2$ bins, respectively.}
\label{fig3}
\end{figure}

\begin{figure}[htbp]
\centering
\begin{tabular}{cc}
\includegraphics[width=70mm]{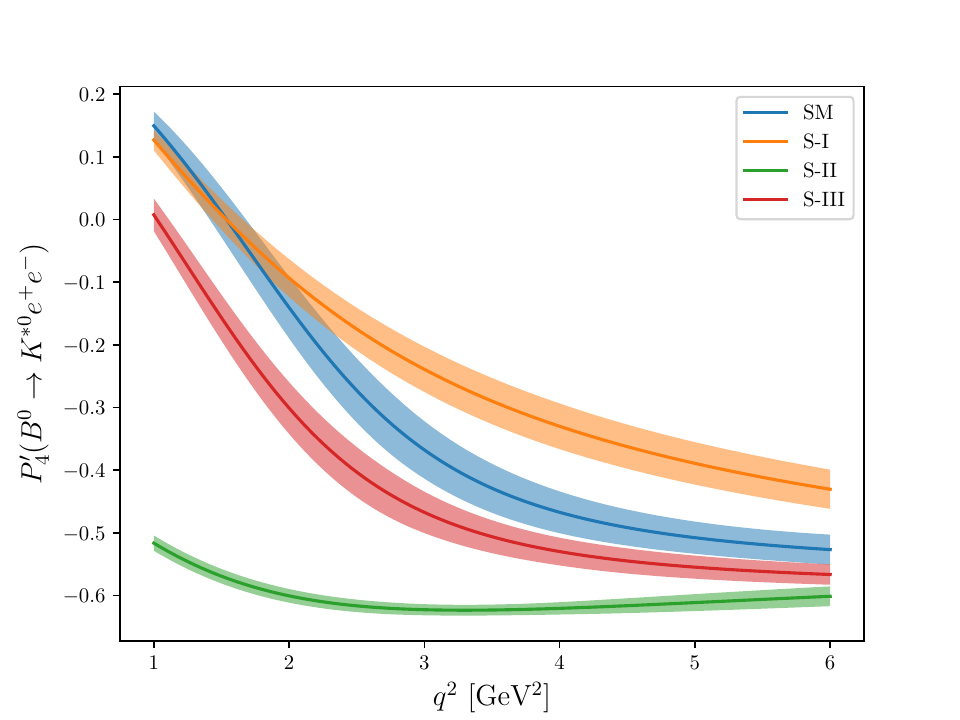} & 
 \includegraphics[width=70mm]{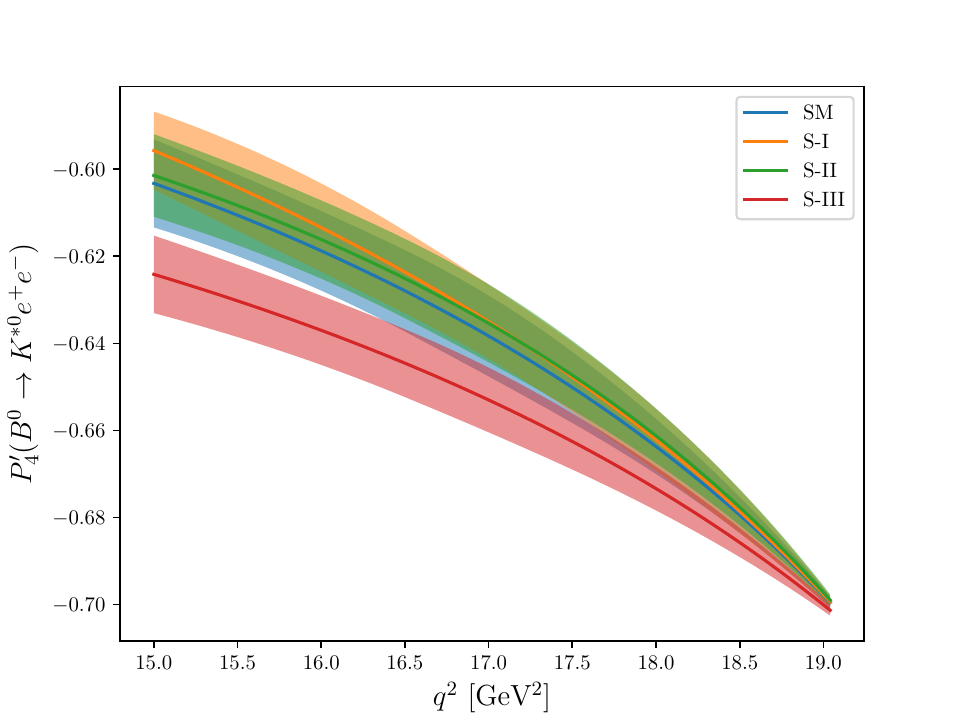}\\
\includegraphics[width=70mm]{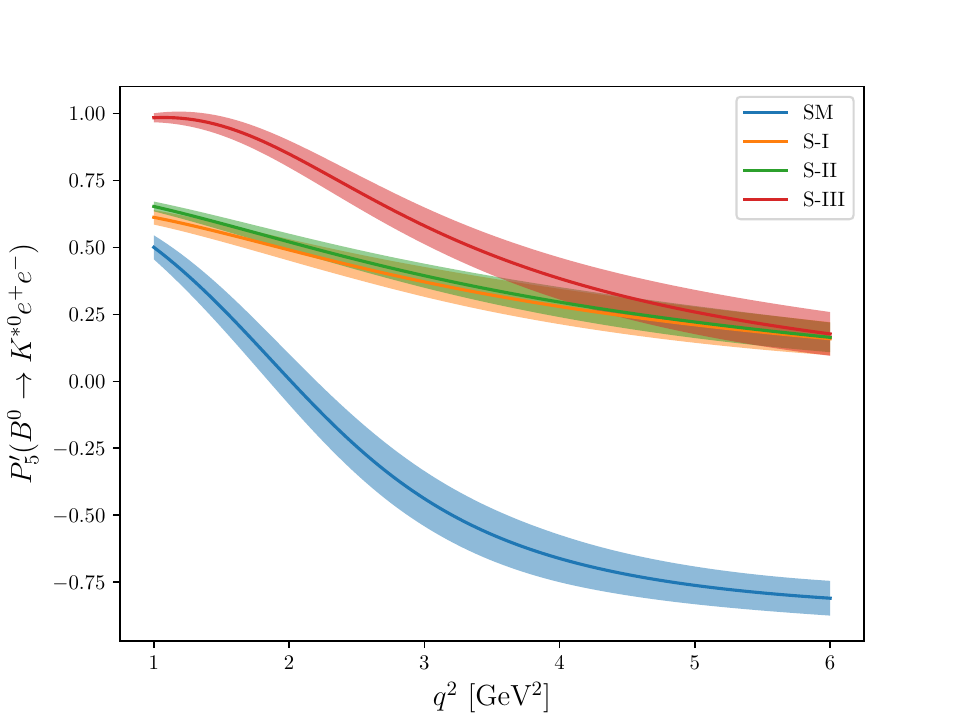} & 
 \includegraphics[width=70mm]{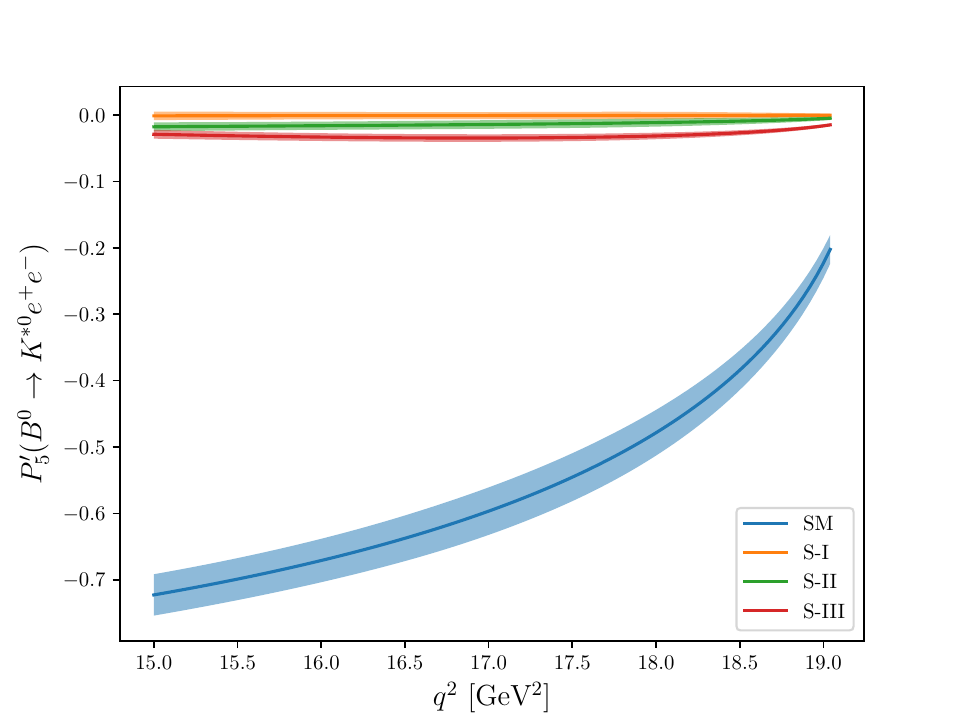}\\
\includegraphics[width=70mm]{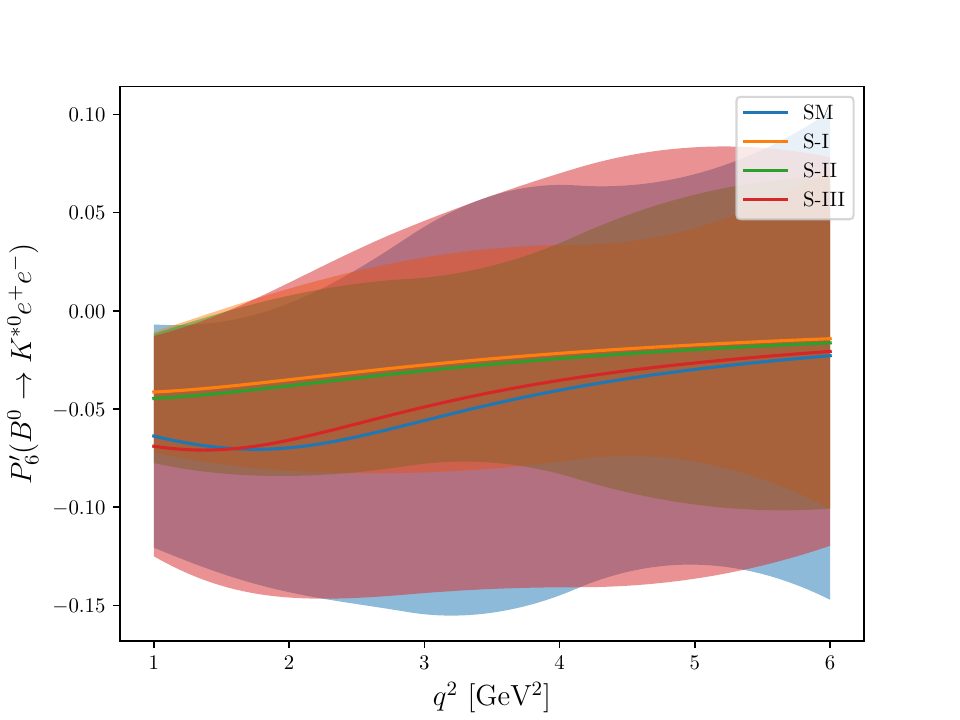} & 
 \includegraphics[width=70mm]{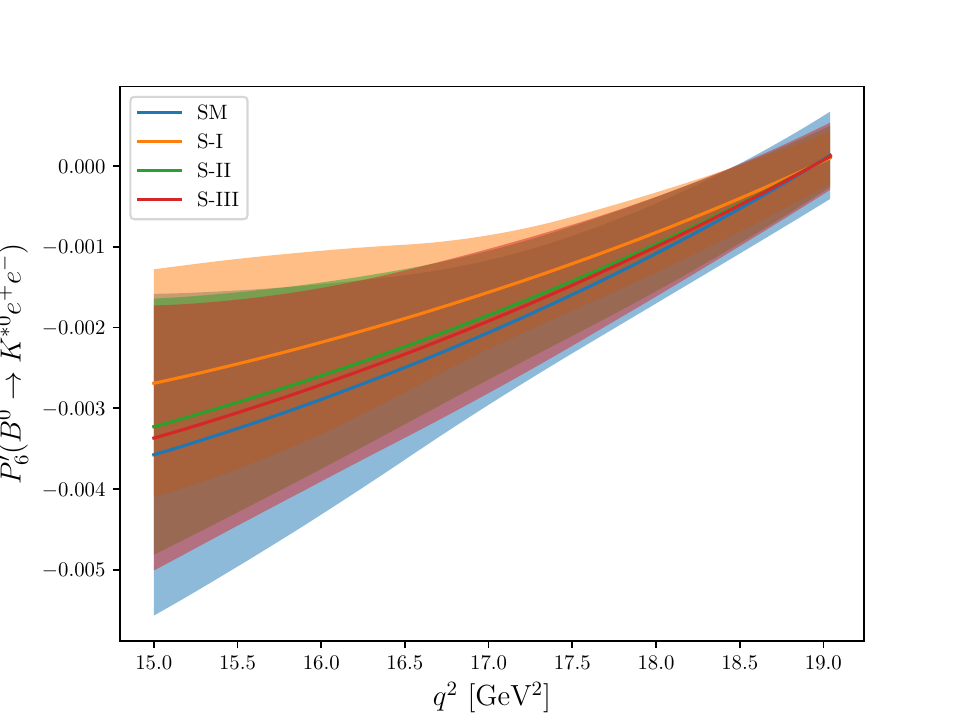}\\
\includegraphics[width=70mm]{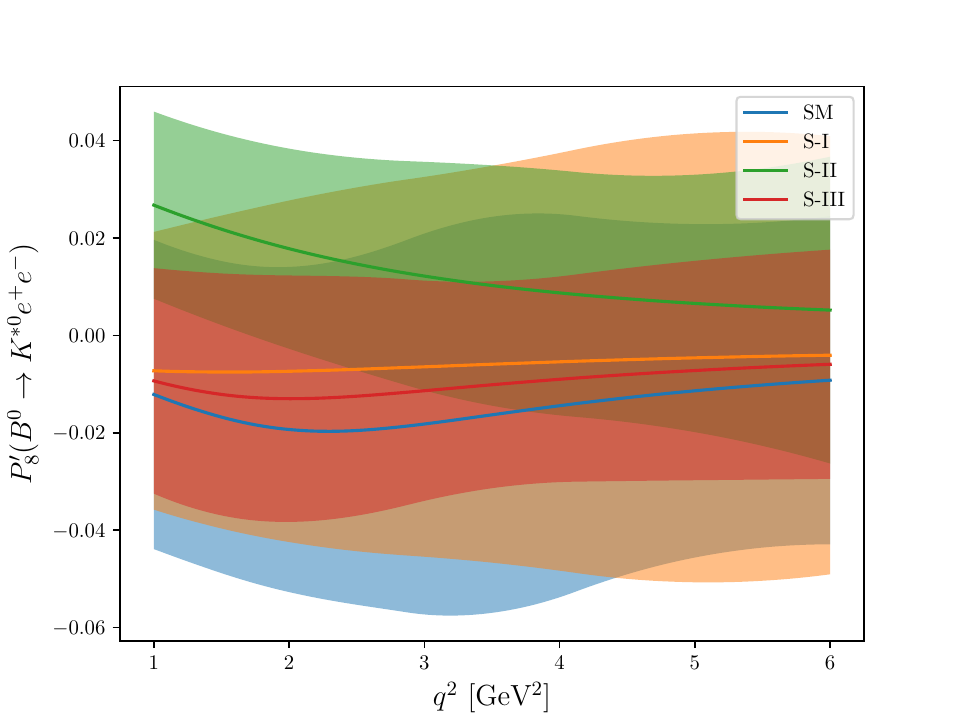} &
 \includegraphics[width=70mm]{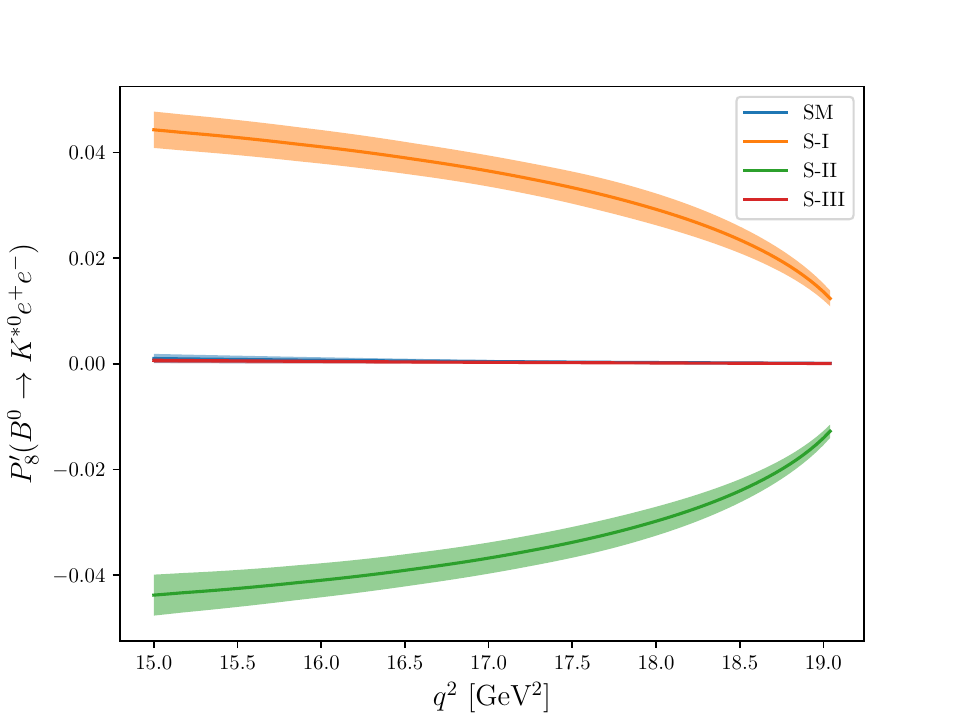}\\
 \end{tabular}
\caption{Plots of $P^{\prime}_{4,5,6,8}(q^2)$ as a function of $q^2$ for the SM and three NP scenarios. The left and right panels correspond to the low ($[1.1,6.0]$ GeV$^2$) and high ($[15,19]$ GeV$^2$) $q^2$ bins respectively.}
\label{fig4}
\end{figure}

Hence we see that neither $A_{FB}$ nor $F_L$ have the power to discriminate between all the three allowed V/A NP solutions. 
Therefore, we now study optimized observables $P_i$ in $B \to K^* e^+ e^-$ decay. In particular, we investigate the distinguishing ability of $P_{1,2,3}$ and $P^{\prime}_{4,5,6,8}$ . We compute the average values of these seven observables for the SM along with  three NP scenarios in two different $q^2$ bins, $q^2\subset [1.1,6.0]$ and $[15,19]$ GeV$^2$. These are listed in Tab~\ref{tabPi}. We also plot these  observables as a function of $q^2$ for the SM and the three  solutions. The $q^2$ plots for $P_{1,2,3}$ and $P^{\prime}_{4,5,6,8}$ are illustrated in Figs.~\ref{fig3} and  \ref{fig4}, respectively.   From these figures and the table, it is apparent that

\begin{table}[ht]
\begin{tabular}{c|c|c|c|c|c}
\hline
Observable & $q^2$ bin & SM & S-I & S-II & S-III  \\
\hline
$P_1$ & $[1.1, 6]$ & $-0.113\pm 0.032$ & $0.507\pm 0.064$ & $-0.627\pm 0.035$ & $-0.291\pm 0.034$ \\
 & $[15,19]$ & $-0.623\pm 0.044$ & $-0.602\pm 0.042$ & $-0.609\pm 0.040$ & $-0.700\pm 0.037$\\
 \hline
$P_2$ & $[1.1, 6]$ & $0.023\pm 0.090$ & $-0.263\pm 0.020$ & $-0.267\pm 0.021$ & $-0.046\pm 0.030$ \\
 & $[15,19]$ & $0.372\pm 0.013$ & $-0.005\pm 0.004$ & $0.002\pm 0.004$ & $0.027\pm 0.004$ \\
 
 \hline
$P_3$  & $[1.1, 6]$ & $0.003\pm 0.008$ & $0.018\pm 0.036$ & $-0.017\pm 0.032$ & $0.002\pm 0.006$ \\
 & $[15,19]$ & $-0.000\pm 0.000$ & $-0.045\pm 0.004$ & $0.045\pm 0.004$ & $-0.000\pm 0.000$  \\
\hline
$P^{\prime}_4$ & $[1.1, 6]$ & $-0.352\pm 0.038$ & $-0.256\pm 0.033$ & $-0.605\pm 0.011$ & $-0.447\pm 0.027$ \\
 & $[15,19]$ & $-0.635\pm 0.008$ & $-0.631\pm 0.008$ & $-0.632\pm 0.008$ & $-0.650\pm 0.008$ \\
 \hline
$P^{\prime}_5$  & $[1.1, 6]$ & $-0.440\pm 0.106$ & $0.336\pm 0.060$ & $0.358\pm 0.045$ & $0.487\pm 0.079$ \\
 & $[15,19]$ & $-0.593\pm 0.036$ & $-0.001\pm 0.005$ & $-0.014\pm 0.006$ & $-0.032\pm 0.005$ \\
 \hline
$P^{\prime}_6$  & $[1.1, 6]$ & $-0.046\pm 0.102$ & $-0.025\pm 0.053$ & $-0.028\pm 0.066$ & $-0.042\pm 0.093$ \\
 & $[15,19]$ & $-0.002\pm 0.001$ & $-0.002\pm 0.001$ & $-0.002\pm 0.001$ & $-0.002\pm 0.001$ \\
 \hline
 $P^{\prime}_8$  & $[1.1, 6]$ & $-0.015\pm 0.035$ & $-0.006\pm 0.032$ & $0.012\pm 0.027$ & $-0.009\pm 0.023$ \\
 & $[15,19]$ & $0.001\pm 0.000$ & $0.036\pm 0.002$ & $-0.036\pm 0.003$ & $0.000\pm 0.000$ \\
 \hline
\end{tabular}
\caption{Average values of $P_{1,2,3}$ and $P^{\prime}_{4,5,6,8}$ in $B\to K^*e^+e^-$ decay for the three V/A NP solutions listed in Table.~\ref{tab3} as well as for the SM.}
\label{tabPi}
\end{table}

\begin{itemize}
\item  The SM prediction of $P_1$ is suppressed in $1-6$ GeV$^2$. However the predicted values for three allowed NP solutions are large and distinct. The reason is following: The observable $P_1$ is very sensitive to the NP WCs $C^{\prime ,e}_9$ and $C^{\prime ,e}_{10}$. In the SM, these NP WCs vanish and this leads to a very small value of $P_1$. On the other hand, the three NP solutions have very large values of $C^{\prime ,e}_9$ and $C^{\prime ,e}_{10}$. Therefore any large values of these NP WCs can lead to a large deviation from the SM prediction. Hence $P_1$  in the low $q^2$ region can discriminate between all three NP solutions, particularly S-I and S-II.  The sign of $P_1$  is opposite for these scenarios. Hence an accurate measurement of  $P_1$ can distinguish between S-I and S-II solutions. In fact, measurement of  $P_1$ with an absolute uncertainty of 0.05 can confirm or rule out S-I and S-II solutions by more than 4$\sigma$. In the high-$q^2$ region, the predictions for all allowed solutions are consistent with the SM. 

\item The observable $P_2$ can be a good discriminant of S-III provided we have handle over its $q^2$ distribution in $[1.1,6.0]$ GeV$^2$ bin. In this bin, $P_2(q^2)$ has a zero crossing at $\sim 3.5$ GeV$^2$ for the SM prediction whereas there is no zero crossing for any of the allowed solutions. Scenarios S-I and S-II predict large negative values for $P_2$, around 
$-0.3$ whereas the S-III predicts relatively smaller negative values. Hence an accurate measurement of $q^2$ distribution of $P_2$  in $[1.1,6.0]$ GeV$^2$ bin can discriminate S-III with other two solutions. In high $q^2$ region, the predictions of $P_2$ for all three solutions are almost the same. These scenarios predict a large deviation from the SM. The SM prediction for $P_2$ is $\sim 0.4$ whereas all three solutions predict values closer to zero. Hence, an accurate measurement of the value of $P_2$ is a smoking-gun signal for the existence of NP in $b \to  s e^+ e^-$ transition as the solution for the current $R_K$ and $R_{K^*}$ anomalies. 

\item The $P_3$ observable in the low-$q^2$ region cannot discriminate between the allowed solutions.  However, in the high $q^2$ region, $P_3$ can uniquely discriminate the three solutions. In particular, the prediction of $P_3$ for S-III in the high $q^2$ is the same as the SM whereas the predictions for S-I and S-II are exactly equal and opposite.

\item The $P^{\prime}_4$ in low-$q^2$ region can only distinguish S-II solution from the other two NP solutions and the SM. In high-$q^2$ region, it has a poor discrimination capability.

\item In the low $q^2$ bin, $P^{\prime}_5$ has a zero crossing at $\sim 2$ GeV$^2$ and has an average negative value in the SM. For all three NP solutions, there is no zero crossing in $P^{\prime}_5$. Further, these scenarios predict a large  positive values. In the high $q^2$ region, the SM predicts a large negative value of $P^{\prime}_5$  whereas NP scenarios predict values close to zero. Thus we see that if we impose  the condition that NP in $b\to se^+e^-$ should simultaneously generate $R_K$ and $R_{K^*}$ within 1$\sigma$ of their measured values, it implies a large deviation in $P^{\prime}_5$ from the SM. This is reflected in the values of pull for the three allowed solutions which are relatively smaller than the other scenarios which fail to explain $R_K$ and $R_{K^*}$ simultaneously. The depletion in pull for these allowed solutions is due to inconsistency between the measured and predicted values of $P^{\prime}_5$.

\item In the both low and high-$q^2$ regions, the NP predictions for $P^{\prime}_6$ for all three scenarios are consistent with the SM. 

\item The $P^{\prime}_8$ in the low-$q^2$ region does not have any discrimination capability. The predicted values for all solutions are consistent with the SM. In the high-$q^2$ region, both S-III and SM predict $P^{\prime}_8$ values close to zero whereas  S-I and S-II predict large positive and negative values, respectively.  
\end{itemize}

From this detailed study of the behavior of the optimized observables $P_i$, we find that both $P_1$ and $P_4^\prime$
at low-$q^2$ have the best capability to discriminate between all the three V/A solutions. The predicted values of $P_1$
are equal and opposite for S-I and S-II and of a much smaller magnitude for S-III. Moreover, each of the predicted
values is appreciably different from the SM prediction. Their magnitudes are quite large $\sim(0.3 - 0.6)$ with a relative 
theoretical uncertainty of about $10\%$. Hence, a measurement of the variable $P_1$, with an experimental uncertainty
of about $0.05$, will not not only confirm the presence of new physics in the $b \to s e^+ e^-$ amplitude but also can
determine the correct WC of the NP operators. In the case $P_4^\prime$, the predictions of all the three solutions have
the same sign but their magnitudes are quite different. The theoretical uncertainty in the predictions is quite low too.
So, $P_4^\prime$ observable also has a good capability to distinguish between the three V/A solutions.

\section{Conclusions}
\label{concl}
In this work,  we intend to analyze $R_{K^{(*)}}$ anomalies by assuming NP only in $b \to se^+e^-$  decay. 
The effects of possible NP are encoded in the WCs of effective operators with different Lorentz structures. 
These WCs are constrained using  all measurements in the $b\rightarrow se^+e^-$ sector along with 
lepton-universality-violating ratios $R_K/R_{K^*}$. We show that scalar/pseudoscalar
NP operators and tensor NP operators can not explain the data in $b \to s e^+ e^-$ sector. 
We consider NP in the form of V/A operators, either one operator at a time or two  similar operators at a time. 
We find that there are several scenarios which can provide 
a good fit to the data. However, there are only three solutions whose predictions of $R_K/R_{K^*}$, 
including $R_{K^*}$ in the in the low-$q^2$ bin ($0.045\,{\rm GeV}^2 \le q^2 \le 1.1 {\rm GeV}^2$), 
match the data well. In order to discriminate between the three allowed V/A solutions, we consider 
several angular observables in the $B \to K^* e^+ e^-$ decay.  
The three solutions predict very different values for the optimized observables $P_1$ and $P_4^\prime$
in the low-$q^2$ bin. Both these observables also have the additional advantage that the theoretical
uncertainties in their predictions are less than $10\%$. Hence a measurement of either of these observables,
to an absolute uncertainty of $0.05$, can lead to a unique identification of one of the solutions.. 
\section*{Acknowledgement}
The work of AKA is partially supported by SERB-India Grant CRG/2020/004576. For partial support, SK acknowledges the IoE-IISc fellowship program.

\end{document}